\documentclass[a4paper]{nature}
\usepackage{epsfig,verbatim}
\usepackage{graphicx}
\usepackage{amsmath}
\usepackage{revsymb}
\usepackage{amssymb}

\allowdisplaybreaks

\mathchardef\ordinarycolon\mathcode`\:
\mathcode`\:=\string"8000
\def\vcentcolon{\mathrel{\mathop\ordinarycolon}}
\begingroup \catcode`\:=\active
  \lowercase{\endgroup
  \let :\vcentcolon
  }
\newcommand{\ket}[1]{|#1\rangle}

\title{Classical signature of quantum
  annealing % \\second in the ``pretending'' series
}

\author{John A. Smolin$^1$ and Graeme Smith$^1$}

\begin{document}
\maketitle
\begin{affiliations}
\item IBM T.J. Watson Research Center, Yorktown Heights, NY 10598, USA
\end{affiliations}

\begin{abstract}

A pair of recent articles\cite{signature,annealing108} 
concluded that the D-Wave One machine actually
operates in the quantum regime, 
% performs quantum annealing, 
rather than performing some classical evolution.  Here we give a
classical model that leads to the same behaviors used in those works to
infer quantum effects.  Thus, the evidence presented does not
demonstrate the presence of quantum effects.
%could equally well be used to demonstrate the classicality of the machine.
% We point out that quantum annealing isn't anealing.....

\end{abstract}
%\maketitle
\section{Introduction}

Adiabatic quantum computation\cite{Farhi2001} has been shown to be
equivalent to the usual circuit model\cite{Equivalent}.  This is only
known to hold for ideal systems without noise.  While there are
effective techniques for fault-tolerance in the circuit
model\cite{ShorFaultTolerance}, it remains unknown whether adiabatic
quantum computation can be made fault-tolerant.

In spite of this, there has been some enthusiasm for implementing
restricted forms of adiabatic quantum computation in very noisy
hardware based on the hope that it would be naturally robust.
Even in the original proposal for quantum
adiabatic computation\cite{Farhi2001} it was suggested it might be a useful
technique for solving optimization problems.  Recent papers about
D-Wave hardware have studied a particular sort of optimization
problem, namely finding the ground state of a set of Ising spins.  These
spins are taken to live on a graph.  The problem instance is
determined by a choice of a graph and either ferromagnetic or
antiferromagnetic interactions between each pair of bits connected by
an edge of the graph.  Finding this ground state is NP-hard if the
graph is arbitrary\cite{Barahona} and efficiently approximable when
the graph is planar\cite{Bansal2009}.  The connectivity of the D-Wave
machine is somewhere in between and it is not known whether the
associated problem is hard.

The D-Wave machine is made of superconducting ``flux''
qubits\cite{Harris2007} (first described in Ref.\cite{Mooij1999}).
Because of the high decoherence rates associated with these flux qubits, it
has been unclear whether the machine is fundamentally quantum or
merely performing a calculation equivalent to that of a classical
stochastic computer.  References\cite{signature,annealing108} attempt
to distinguish between these possibilities by proposing tests for
quantumness that the D-Wave machine passes but a purely classical
computer should fail. This letter presents a classical model that
passes the tests, exhibiting all the supposedly quantum behaviors.

%by by drawing a line between
%classical simulated annealing and what is referred to as ``quantum
%annealing.''
%The purpose of this letter is to demonstrate that this
%reasoning is faulty because it excludes the possibility of something
%other than either classical or quantum annealing.  We give a purely
%classical model that reproduces all the behaviors taken to be
%signatures of the quantum.

\section{The claims}
\subsection{Eight-spin signature of quantum annealing}

\begin{figure}[htbf]
\centering{\includegraphics[height=2in]{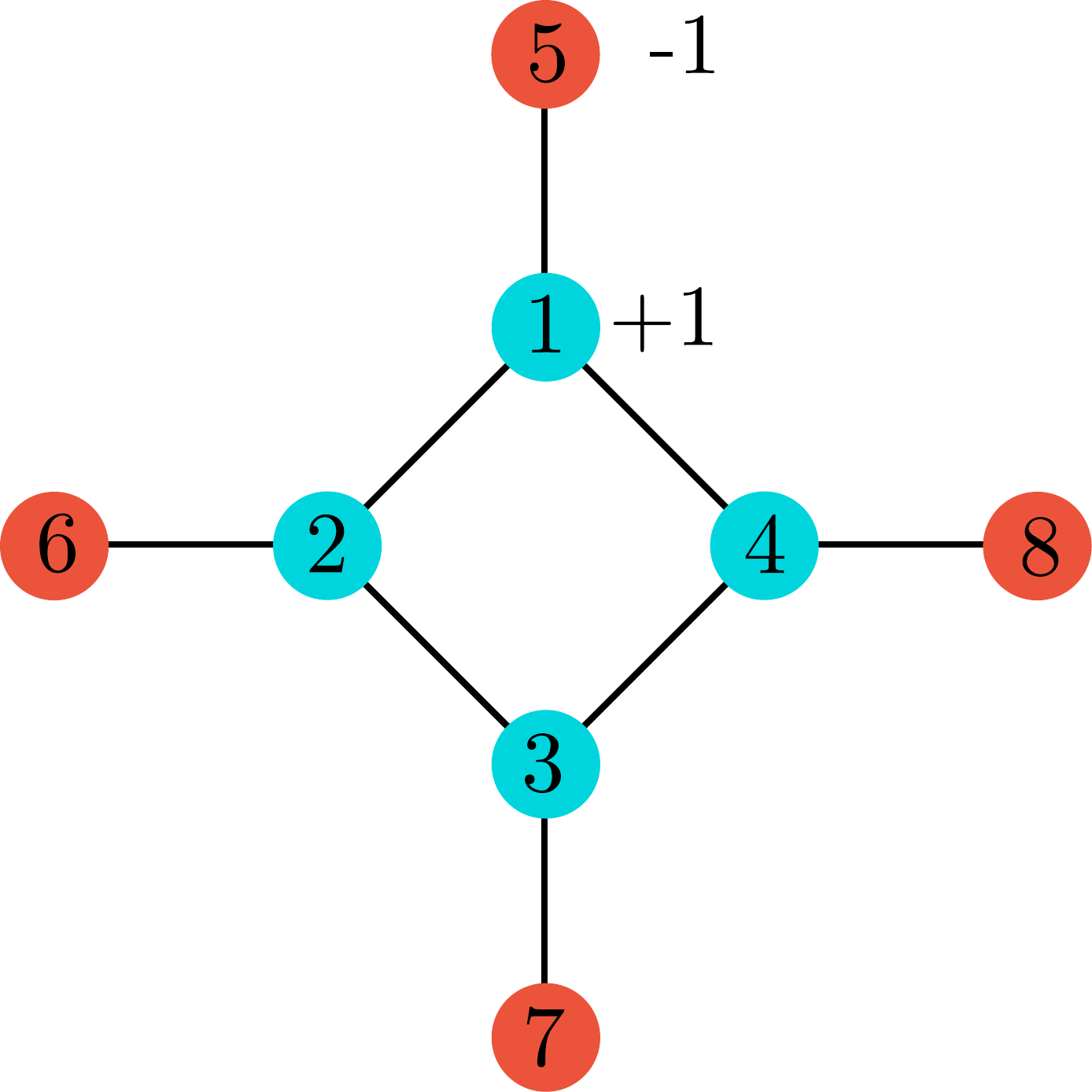}}
\caption{The eight spin graph from Reference\protect\cite{signature}.  Each spin is
coupled to its neighbors along the graph with a $+1$ coupling such
that the spins like to be aligned.  In addition, the inner ``core'' spins
have a magnetic field applied in the $+z$ direction while the outer
``ancillae'' spins have a field applied in the $-z$ direction.
\label{fig:eight}}
\end{figure}

In\cite{signature}, a system of eight spins as shown in Fig.\ \ref{fig:eight}
is analyzed.  It is shown that the ground state of the system is
17-fold degenerate, comprising the states
\begin{equation}
\{\ket{\!\!\uparrow\uparrow\uparrow\uparrow\downarrow\downarrow\downarrow\downarrow},\ldots,
\ket{\!\!\uparrow\uparrow\uparrow\uparrow\downarrow\uparrow\downarrow\downarrow},
\ldots,
\ket{\!\!\uparrow\uparrow\uparrow\uparrow\uparrow\uparrow\uparrow\uparrow}\}\ \mathrm{and}\ 
\ket{\!\!\downarrow\downarrow\downarrow\downarrow\downarrow\downarrow\downarrow\downarrow}\ .
\end{equation}
The probability of finding the isolated (all down) state $p_s$ is compared
to the average probability of states from the 16-fold ``cluster'' of states,
$p_C$.  It is computed that classical simulated annealing finds an 
enhancement of $p_s$, \emph{i.e.} $p_s > p_C$ while quantum
annealing both in simulation and running on the D-Wave machine finds
a suppression $p_s < p_C$.

\subsection{Annealing with 108 spins}

In Ref.\cite{annealing108} 108 spins in the D-Wave computer are employed.
Their connectivity is displayed in Figure 1 of that paper.  Each
connection is given a coupling of $\pm 1$ at random.  1000 random
cases are studied, and for each case the machine is run 1000
times.\footnote{They also look at what happens with fewer than 108
  spins, so the total number of experiments they performed is actually
  much larger.}  The ground state energies calculated are compared to
the correct answer, which can be found for cases of this
size\cite{spinglassserver}.  A histogram of the probabilities of
successfully finding the correct answer is shown.  The histogram has a
bimodal distribution, with significant peaks at probability zero and
probability one.  The cases where the machine never find the answer
are ``difficult'' cases and the ones where it always finds it are
``easy'' cases.  By comparison, classical simulated annealing shows a
unimodal distribution with no, or nearly no, cases being either hard
or easy.  See Figure 2 in Ref.\cite{annealing108}

\section{Quantum annealing is not annealing}
Is it surprising that results of the D-Wave experiments differ greatly
from classical simulated annealing?  And should this be considered evidence
that the machine is quantum or more powerful than classical computation in 
some way?  We argue here that the answer to these questions is ``no.''

What is called ``quantum annealing'' is often compared to classical
simulated annealing\cite{SimulatedAnnealing}, or to physical annealing
itself.  Though both quantum annealing and simulated annealing are
used to find lowest-energy configurations, one is not the quantum
generalization of the other.  More properly, quantum annealing should
be considered quantum adiabatic ground state dragging. 

Simulated annealing proceeds by choosing a random starting state
and/or a high temperature, then evolving the system according to a set
of rules (usually respecting detailed balance) while reducing the
simulated temperature.  This process tends to reduce the energy during
its evolution but can, of course, become stuck in local energy minima,
rather than finding the global ground state.

Quantum annealing, on the other hand, involves no randomness or
temperature, at least in the ideal.  Rather, it is a particular type
of adiabatic evolution.  Two Hamiltonians are considered: The one for
which one desires to know the ground state $H_f$, and one which is
simple enough that cooling to its ground state is easy, $H_i$.  At the
start of the process, the system is initialized to the ground state of
$H_i$ by turning on $H_i$ and turning off $H_f$ and waiting for
thermal equilibration.  Then $H_i$ is gradually turned off while $H_f$
is gradually turned on.  If this is done slowly enough, the system
will at all times remain in the ground state of the overall
Hamiltonian.\footnote{If the ground state is degenerate at some point
  during the process then this may not be true.}  Then, at the end of
this process the system will be in the ground state of
$H_f$.\footnote{Of course, ``slowly enough,'' depends on the gap
  between the ground state and other nearby energy eigenstates.  If
  the system is frustrated, there are many such states and the
  evolution must proceed exponentially slowly, just as frustration
  hinders classical simulated annealing.  Though it has often been
  suggested that quantum annealing is a panacea, whether it can
  outperform classical simulated annealing in such cases is unknown.}
It remains to measure the ground state.  For the problems considered
in Refs.\cite{signature,annealing108}, the ground state is typically known
to be diagonal in the $z$ basis.  Thus, a measurement need not
introduce any randomness.

Since classical simulated annealing is intrinsically random and
``quantum annealing'' is not, the differences reported in
Refs.\cite{signature,annealing108} are not surprising.  For the
eight-bit suppression of finding the isolated state, two things could
have happened:  Either the ideal machine would find the isolated state
always, or never.  It happens that, due to the structure of the state,
the ideal outcome is ``never,'' which is certainly a suppression.  The
bimodal distribution found for the 108-bit computations is also
just what one would expect: In the perfect case of no noise either
the calculation gets the correct answer or it does not.  The outcome
is deterministic so there should be exactly two peaks, at probability
of success zero and one.  Classical annealing, which begins from a
random state on each run, is not expected to succeed with probability
one, even for cases where the system is not frustrated.

The bimodality of the D-Wave results, in contrast to the unimodality
of simulated annealing, can be seen as evidence not of the machine's
quantumness, but merely of its greater reproducibility among runs
using the same coupling constants, due to its lack of any explicit
randomization.  The simulated annealing algorithm, by contrast uses
different random numbers each time, so naturally exhibits more
variablity in behavior when run repeatedly on the same set of coupling
constants, leading to a unimodal historgram. Indeed if the same random
numbers were used each time for simulated annealing, the histogram
would be perfectly bimodal.  To remove the confounding influence of
explicit randomization, we need to consider more carefully what would
be a proper classical analog of quantum annealing.

If, as we have argued, classical simulated annealing is not the correct
classical analog of quantum annealing, what is?  The natural answer
is to classically transform a potential landscape slowly enough that
the system remains at all times in the lowest energy state.  
In the next section we give a model classical system and, by running
it as an adiabatic lowest-energy configuration finder, demonstrate that
it exhibits the same computational behavior interpreted as a quantum
signature in Refs.\cite{signature,annealing108}.

\section{The model}

The flux qubits in the D-Wave machine decohere in a time considerably
shorter than the time adiabatic evolution experiment runs.  The
decoherence times are stated to be on the order of tens of nanoseconds
while the adiabatic runtime is 5-20 microseconds\cite{annealing108}.
For this reason, one would expect that no quantum coherences should
exist.  We therefore model the qubits as $n$ classical spins
(``compass needles''), each with an angle $\theta_i$ and coupled to
each other with coupling $J_{ij}=0,\pm 1$.  Each spin may also be acted
on by external magnetic fields $h_i$ in the $z$-direction as well as
an overall field in the $x$-direction $B_X$.  The potential energy
function is then given by
\begin{equation}
V_\mathrm{Ising}=-\sum_i \cos(\theta_i) h_i - \frac{1}{2}\sum_{i\ne j} \cos(\theta_i)\cos(\theta_j) J_{ij}
\label{eq:vising}
\end{equation}
and
\begin{equation}
V_\mathrm{trans}=-\sum_i \sin(\theta_i) B_x \ .
\label{eq:vtrans}
\end{equation}
Compare these to Equations (1) and (2) in Ref.\cite{signature}

The adiabatic computation is performed (or simulated) by running
the dynamics while gradually changing these potentials from 
$V_\mathrm{trans}$ to $V_\mathrm{Ising}$ over a time $T$ 
according to
\begin{equation}
V(t)=A(t) V_\mathrm{trans}+ B(t) V_\mathrm{Ising}
\label{eq:vtotal}
\end{equation}
with $A(0)=B(T)=1$ and $A(T)=B(0)=0$.
The equations of motion are simply:
\begin{equation}
\frac{d}{dt}\theta_i=\dot{\theta}_i\ \mathrm{and}\ \frac{d}{dt}\dot{\theta}_i=\frac{dV}{d\theta_i} \ .
\label{eq:motion}
\end{equation}
It is straightforward to integrate this system of ordinary
differential equations.

\section{Results}
\subsection{Eight spin model}
\begin{figure}[htbf]
\centering{\includegraphics[height=2.5in]{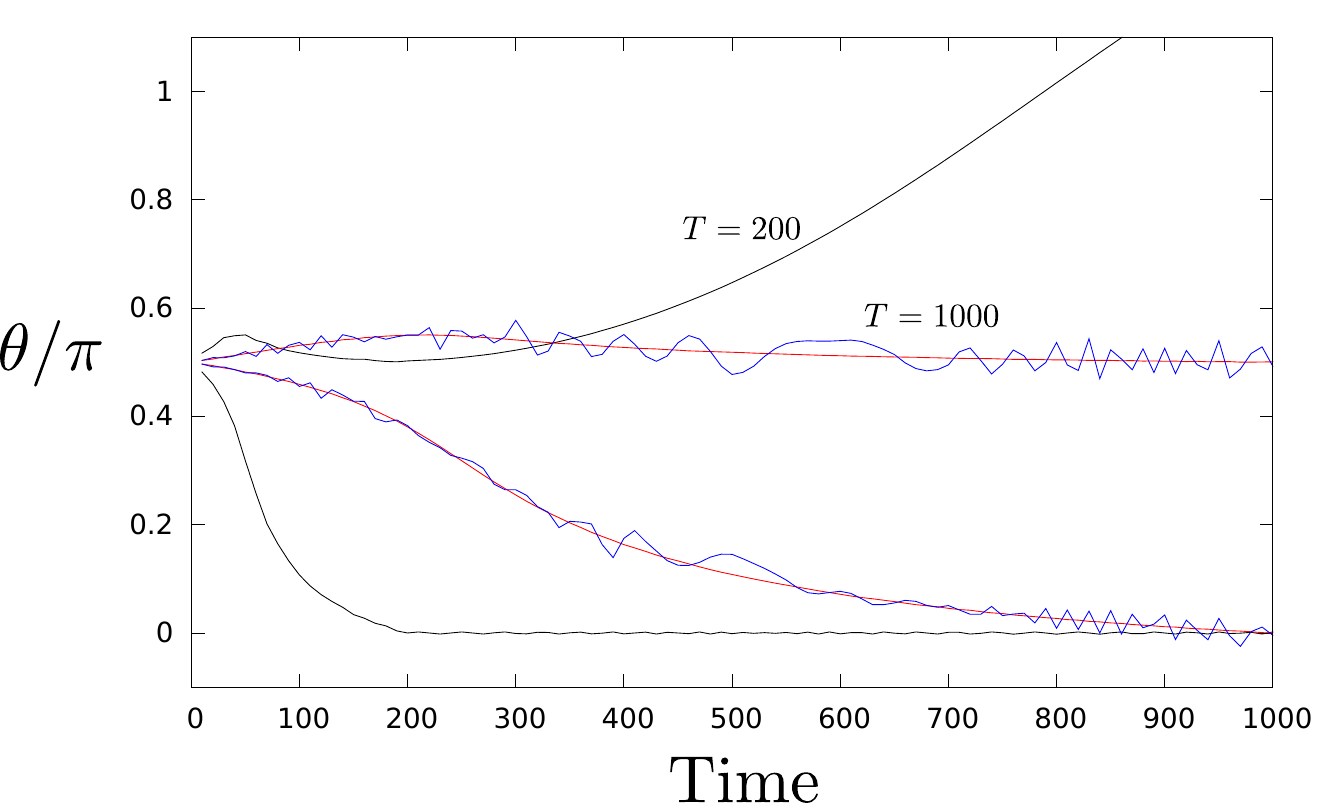}}
\caption{Results for eight spin model.  Three runs are shown.  In each
  case the angle $\theta$ of a representative core spin and ancillae spin are
  shown as a function of time.  The core spins are all driven to
  $\theta=0$ while the ancillae spins are ideally driven to
  $\theta=\Pi/2$.  1) The red lines show a case with no noise and with
  adiabatic drag time $T=1000$.  2) The blue lines have added noise
  again with $T=1000$. The noise was simulated by applying random
  kicks uniformly distributed between $\pm 0.02$ to the angular
  velocities $\dot{\theta}$ of the spins at $t=10,20, \ldots$.  3) The
  black lines have no noise but $T=200$ (the evolution continues after
  the adiabatic drag is complete.  It is easily seen the ancillae spin
  has been driven too fast and winds up with some kinetic energy (the
  slope is not zero for $t>T$).
  \label{fig:8spinplot}}
\end{figure}

The simulated adiabatic dragging time $T$ needs to be long compared to
the fundamental timescales of the system.  Since units have been
omitted in Eqs.\ (\ref{eq:vising}-\ref{eq:motion}), these are of order
unity.  Using $T=1000$ produces good results as shown in the next
section.  Fig.\ \ref{fig:8spinplot} shows the results of simulating
the classical model of the eight spins from Ref.\cite{signature}
When the total adiabatic dragging time $T$ is long enough, the
dynamics lead to a stationary state (red lines).  The core spin is
driven to $\theta=0$, corresponding to $\ket{\!\uparrow}$, and the
ancillae spin to $\theta=\frac{\pi}{2}$, corresponding to
$\ket{\!\!\rightarrow}$. All other core and ancillae states behave
identically.  The final state
$\ket{\!\!\uparrow\uparrow\uparrow\uparrow\rightarrow\rightarrow\rightarrow\rightarrow}$
lies in the space spanned by the 16 cluster states.  Within this space
the ancillae spins are free to rotate as they contribute nothing to
the energy provided the core states are all $\ket{\!\!\uparrow}$.
Since there is nothing to break the symmetry of the initial all
$\ket{\!\!\rightarrow}$ configuration, this is the final state they
choose. This behavior is robust to added noise (blue lines).  However,
when the dragging time $T$ is too short, the long-time behavior is not
a stationary state (black lines).

\subsection{108 spins}

For this case we programmed 108 spins with the same connectivity 
as run in Ref.\cite{annealing108}  Choosing 500 random
cases of $\pm$ for the couplings and running 
%the VM
our classical compass model simulation
with no noise and a dragging time $T=1000$ results in a perfect
bimodal distribution.  For 249 cases the optimal answer was found
always, and for 251 it was found never (since the machine is
deterministic it is only necessary to run each case once).  It seems
to be a coincidence that for problems of this size the number of easy
and hard cases is almost equal.
 
\begin{figure}[htbf]
\centering{\includegraphics[height=3in]{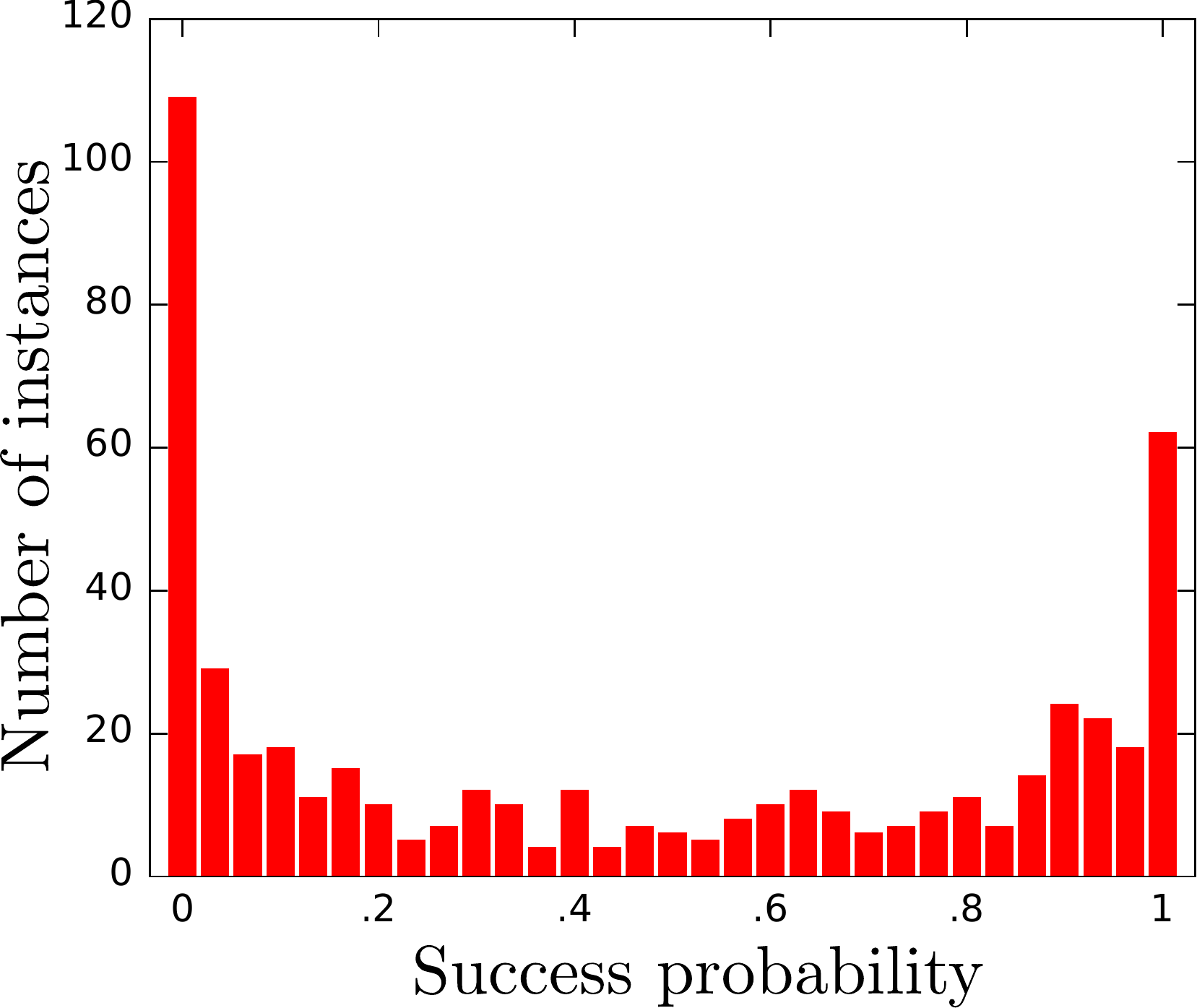}}
\caption{Results of 500 spin glass instances, each run 30 times on a
  noisy simulation of the classical compass model with $T=1000$.  
   A bimodal distribution is
  observed, with a clear separation between easy and hard instances
  with high and low success probabilities respectively.  The noise
  applied was random kicks to each $\dot{\theta}$ at $t=10,20,\ldots,1000$
 uniformly chosen between $\pm 0.0015$.  
\label{fig:bimodal}}
\end{figure}

We show in Fig.\ \ref{fig:bimodal} the results of running a simulation
of the classical compass model with with noise.  For the same 500 sets
of couplings, 30 noisy runs were performed and a histogram of success
probabilities results.  The bimodal distribution is maintained.
Compare to Fig.\ 2 in Ref.\cite{annealing108}  The qualitative nature
of the bimodal signatures found is insensitive to the details of the
noise so more realistic noise models would give similar behavior.
Note also that the noise seems to help find the ground state, as in
401 of the 500 cases the ground state was found at least once in 30
trials.  This suggests that noisy adiabatic dragging picks up some of
the benefits of classical simulated annealing, avoiding being
``stuck'' with either always or never finding the ground state.

\section{Conclusions}

We have argued that quantum annealing and simulated annealing are very
different procedures.  The deterministic nature of quantum annealing
leads to rather different behaviors than the random processes of
simulated annealing.  However, other deterministic procedures can also
lead to behavior very similar to that observed in the D-Wave device.
Our classical model reproduces all the claimed signatures of quantum
annealing.  We recommend using the term ``ground-state adiabatic dragging'' 
or simply ``adiabatic computation'' for such nonrandom processes.

Note that in Ref.\cite{manufactured}, under highly diabatic
conditions, evidence was shown that the D-Wave device exhibits quantum
tunneling.  We emphasize that this does not constitute
evidence for an essential use of quantum effects during adiabatic
dragging and indeed an effectively classical model may well capture
the physics and computational behavior seen in
Refs.\cite{signature,annealing108}.  Tunneling is a possible
additional path for a system to anneal into a lower energy state
without needing enough energy to cross a potential barrier.  If the
decoherence is such that each spin is projected into a local energy
eigenstate, then in the adiabatic limit, where each spin is at all
times in its local ground state, tunneling locally can be of no
assistance.  The tunneling in Ref.\cite{manufactured} is indeed only
shown locally.

Furthermore, there is nothing preventing the implementation of our
simulated compass model in hardware.  It would be possible to build an
analog classical machine that could simulate it very quickly, but it
would be simpler use digital programmable array with one processing
core per simulated spin.  Since each spin requires knowledge of at
most six of its neighbors along the connectivity graph, the algorithm
can be easily parallelized.  A 108 core computer specialized to
running our algorithm could easily run hundreds or thousands of times
faster than simulating it on a desktop computer, and could be built
for modest cost using off-the-shelf components.  Similarly, \emph{any}
classical physics can be efficiently simulated on a classical machine.

This is not to suggest that simulating classical physics directly on
a classical computer is a good way to solve
optimization problems.  Classical simulated
annealing\cite{SimulatedAnnealing} and other heuristic techniques have
been extremely successful, and can also be very fast on
special-purpose hardware.  It is shown in Ref.\cite{annealing108} that
simulated annealing is competitive with the D-Wave One machine even on
desktop CPUs.  Stronger evidence of both quantumness and a resulting
algorithmic speedup is needed before quantum adiabatic computers will
have proved their worth.

\vspace{1 in}

\bibliographystyle{unsrt}

\bibliography{compass}

\begin{thebibliography}{10}

\bibitem{signature}
Sergio Boixo, Tameem Albash, Federico~M. Spedalieri, Nicholas Chancellor, and
  Daniel~A. Lidar.
\newblock Experimental signature of programmable quantum annealing.
\newblock ar{X}iv:1212.1739, 2012.

\bibitem{annealing108}
Sergio Boixo, Troels~F. Ronnow, Sergei~V. Isakov, Zhihui Wang, David Wecker,
  Daniel~A. Lidar, John~M. Martinis, and Matthias Troyer.
\newblock Quantum annealing with more than one hundred qubits.
\newblock ar{X}iv:1304.4595, 2013.

\bibitem{Farhi2001}
Edward Farhi, Jeffrey Goldstone, Sam Gutmann, Joshua Lapan, Andrew Lundgren,
  and Daniel Preda.
\newblock A quantum adiabatic evolution algorithm applied to random instances
  of an np-complete problem.
\newblock {\em Science}, 292(5516):472--475, 2001.

\bibitem{Equivalent}
Dorit Aharonov, Wim van Dam, Julia Kempe, Zeph Landau, Seth Lloyd, and Oded
  Regev.
\newblock Adiabatic quantum computation is equivalent to standard quantum
  computation.
\newblock {\em SIAM J. Comput.}, 37(1):166--194, April 2007.

\bibitem{ShorFaultTolerance}
Peter~W. Shor.
\newblock Fault-tolerant quantum computation.
\newblock In {\em Proceedings of the 37th Symposium on Foundations of
  Computing}, FOCS 1996, pages 56--65, 1996.

\bibitem{Barahona}
F.~Barahona.
\newblock On the computational complexity of ising spin glass models.
\newblock {\em Journal of Physics A: Mathematical and General}, 15(10):3241,
  1982.

\bibitem{Bansal2009}
Nikhil Bansal, Sergey Bravyi, and Barbara~M. Terhal.
\newblock Classical approximation schemes for the ground-state energy of
  quantum and classical ising spin hamiltonians on planar graphs.
\newblock {\em Quantum Info. Comput.}, 9(7):701--720, July 2009.

\bibitem{Harris2007}
R.~Harris, A.~J. Berkley, M.~W. Johnson, P.~Bunyk, S.~Govorkov, M.~C. Thom,
  S.~Uchaikin, A.~B. Wilson, J.~Chung, E.~Holtham, J.~D. Biamonte, A.~Yu.
  Smirnov, M.~H.~S. Amin, and Alec Maassen van~den Brink.
\newblock Sign- and magnitude-tunable coupler for superconducting flux qubits.
\newblock {\em Phys. Rev. Lett.}, 98:177001, Apr 2007.

\bibitem{Mooij1999}
J.~E. Mooij, T.~P. Orlando, L.~Levitov, Lin Tian, Caspar~H. van~der Wal, and
  Seth Lloyd.
\newblock Josephson persistent-current qubit.
\newblock {\em Science}, 285(5430):1036--1039, 1999.

\bibitem{spinglassserver}
Spin glass server.
\newblock http://www.informatik.uni-koeln.de/spinglass/.

\bibitem{SimulatedAnnealing}
S.~Kirkpatrick, C.~D. Gelatt, and M.~P. Vecchi.
\newblock Optimization by simulated annealing.
\newblock {\em Science}, 220(4598):671--680, 1983.

\bibitem{manufactured}
M.W. Johnson, M.H.S. Amin, S.~Gildert, T.~Lanting, F.~Hamze, N.~Dickson,
  R.~Harris, A.J. Berkley, J.~Johansson, P.~Bunyk, et~al.
\newblock Quantum annealing with manufactured spins.
\newblock {\em Nature}, 473(7346):194--198, 2011.

\end{thebibliography}

% \textbf{Supplementary Information} is available in the online
% version of the paper.

\textbf{Acknowledgments}
The authors thank Charles Bennett, Jay Gambetta, Mark Ritter, and 
Mattias Steffen for helpful comments on our manuscript.

%\textbf{Author Information} Reprints and permissions information is
%available at www.nature.com/reprints. The authors declare no competing
%financial interests.  Readers are welcome to comment on the online
%version of the paper. Correspondence and requests for materials should
%be addressed to J.A.S. (smolin@alum.mit.edu).

\end{document}